\documentclass[twocolumn,aps,prb,showpacs,preprintnumbers,amsmath,amssymb]{revtex4-1}
\usepackage{natbib}
\usepackage{epsfig}
\usepackage{graphicx}
\usepackage{color} 
\usepackage{amsfonts}

\newcommand{\ct}{\cite}
\newcommand{\bi}{\bibitem}
\newcommand{\be}{\begin{equation}}
\newcommand{\ee}{\end{equation}}
\newcommand{\ba}{\begin{eqnarray}}
\newcommand{\ea}{\end{eqnarray}}

\newcommand{\ket}[1]{|#1\rangle}
\newcommand{\de}{\delta}

\begin{document}
\title{Survival probability of an edge Majorana  in a 1-D p-wave superconducting chain under sudden quenching of parameters}
\author{Atanu Rajak$^1$}
\author{Amit Dutta$^2$}
\affiliation{$^1$CMP Division, Saha Institute of Nuclear Physics, 1/AF Bidhannagar, Kolkata 700 064, India}
\affiliation{$^2$Department of Physics, Indian Institute of Technology Kanpur, Kanpur 208016, India}
\begin{abstract}
We consider the 
temporal evolution of a zero energy edge Majorana of  a  spinless $p$-wave superconducting chain  following a 
sudden change of a parameter of the Hamiltonian.    Starting from one of the topological phases
that has an edge Majorana, the system is suddenly  driven to the  other topological phase or
to the (topologically) trivial phases and also 
to the quantum critical points (QCPs) separating these phases. The survival probability of the  initial edge Majorana as a function of time is studied following 
the quench. Interestingly when the chain is quenched to the QCP, we find a nearly perfect oscillations of the survival probability, indicating that the Majorana 
travels back and forth between two ends,  with a time period that scales with the system size. We also generalize to the situation when there is a next-nearest-neighbor
 hopping in superconducting chain and there
resulting in  a pair of edge Majorana at the each end of the chain in the topological phase. We show that the frequency of oscillation of the survival probability 
gets doubled in this case. We also perform an instantaneous quenching the Hamiltonian (with two Majorana modes at each end of the chain) to an another Hamiltonian which  has  only one Majorana 
mode in equilibrium; the MSP shows oscillations as a function of time with a noticeable decay in the amplitude. On the other hand for a quenching
which is reverse to the previous one,   the MSP decays rapidly and stays close to zero  
with fluctuations in amplitude.
\end{abstract}
\maketitle
\date{today}
\section{Introduction}
\label{I}
The topological properties of a $p$-wave spin-less superconductor, introduced by Kitaev \ct{kitaev01} has been  studied from different points of 
views in recent years \ct{fulga11,sau12,lutchyn11,degottardi11,degottardi13,thakurathi13,wdegottardi13}. (For a review see \ct{alicea12}). A spinless $p$-wave 
superconductor is a topological system which has edge Majorana modes as midgap excitations that are guaranteed by particle-hole symmetry of the system. The superconducting 
pairing term of such a Hamiltonian actually induces the zero energy excitations in the system.
The distinct phases of this system are characterized  by  the presence (or absence) of the zero energy  Majorana modes at the ends of a long and open chain. 
The phases those have zero energy Majorana modes are called topologically non-trivial and others are topologically trivial. 
The number of these modes are topological invariant;  this value does not change until the system crosses the phase boundary  
between the topological and the trivial phases as happens in the case of a topological insulator \ct{hasan10,qi11}.
It has been proposed that the proximity effect between the surface states of a strong topological insulator and $ s$-wave superconductor can generate a two dimensional state strongly resembling a $p$-wave superconductor which  can hosts Majorana states at the vortices\ct{fu08}.

Recent experiments have also been able to detect the signature of these Majorana modes 
in the zero-bias transport properties of nanowires coupled to superconductors~\ct{Kouwenhoven12,deng12,das12,chang13}. 
Additionally, it has also been demonstrated experimentally that these Majorana modes can be hybridized and be made to appear or disappear by tuning the 
chemical potential of a similar system across a topological phase transition~\ct{Finck13}. There are also some discrepancies in experimental 
results with the theoretical predictions of Majorana fermions which has been discussed in Ref. \onlinecite{rainis13}.
We note that there are several recent studies on the 
possible decoherence of a Majorana qubit \ct{budich12,schmidt12} and the
possibility of dynamical generation of Majorana Fermions \ct{thakurathi13,perfetto13}; the sudden quench of a 1-d Majorana chain has also been studied through its 
topological signature in the entanglement spectrum \ct{chung14}.

In parallel, there are a plethora of recent studies of the non-equilibrium  dynamics of quantum many body systems driven across  quantum critical points\ct{polkovnikov10,dutta10,dziarmaga10}. The failure of achieving a  perfectly adiabatic evolution 
close to a  quantum critical point (QCP), where the relaxation time is very large, produce excitations in the system which leads to non-zero defect density in the final state that the system reaches
following the  quenching. This defect density scales with the 
rate of quenching which is usually determined by the  Kibble-Zurek (KZ) scaling relation\ct{zurek05,polkovnikov05,damski05} given in terms of
the rate of quenching and some of the critical exponents associated with the QCP across which the system is driven. On the other hand, for a sudden quench  across a QCP, the scaling of the defect density generated in the final state is given in terms of the magnitude of the change
of the quenched parameter and the critical exponents \ct{grandi10,mukherjee11}. Recently, the studies of the connection between 
quenching dynamics and the topological order \ct{dimitris09,rahmani10,halasz13} and also the dynamics of an edge state existing in the topological phase\ct{bermudez09} have attracted the attention of the scientists . It has been established that following a  slow quenching across
a QCP, an edge  Majorana of 1-D $p$-wave superconductor shows different scaling from that predicted by the Kibble-Zurek scaling indicating that  the quenching dynamics  of an edge state is non-universal and depends on the topology of the system\ct{bermudez10}.  In a similar spirit, the 
dynamics of an edge state of a  topological insulator in a  ribbon geometry has been explored after the system  is suddenly quenched to the QCP (between
the topological and trivial insulator phases) and the signature of a surviving 
spin-Hall current that oscillates in time has been  observed\ct{patel13}.  We note that very recently the possibility of thermalization of non-local
topological order has been studied  by tracking the time evolution of the string
correlations in a spin-1 chain in its Haldane phase  following a sudden quench \ct{mazza13}.

In this paper, we consider a spinless $p$-wave superconducting chain of finite but long system size and study the dynamics of a zero energy edge Majorana after suddenly 
quenching the system from one topologically non-trivial phase to another (or to the topologically trivial phase) and also  to the QCP separating these phases.  Following the quench, the edge Majorana gets coupled to the
bulk modes and hence is expected to decohere. We address the question whether this happens for all quenching schemes. Analyzing the Majorana survival probability (MSP), we here demonstrate that
 there could be  situations when the edge Majorana survives and oscillates between two edges of the chain.

The paper is organized in the following manner: in the next section (Sec.\ref{II}) we discuss the model Hamiltonian discussing the equilibrium phase diagram with different topological and non-topological 
phases. In Sec. \ref{III} we shall define  the quantity which we call the  MSP
and  measure  this after a rapid quench of a parameter of the superconducting chain. In subsequent subsections, we consider different paths for quenching in the 
phase diagram
and present our numerical results. We extend the results to  1-D chain to study  of $p$-wave superconductor with next nearest neighbor hopping in the Sec. \ref{IV}.
We also calculate the MSP in Sec.~\ref{V} when the system is quenched between a phase with two Majorana modes to a phase with one Majorana mode.
 Finally,  our results are summarized  in Sec. \ref{VI}.

\section{Model}
\label{II}
\begin{figure}[ht]
\begin{center}
\includegraphics[height=2.4in]{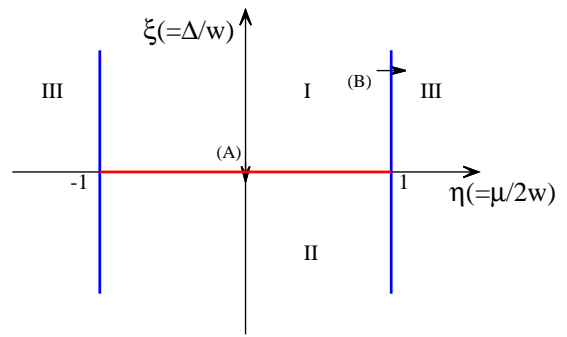}
\end{center}
\caption{(Color online) Phase diagram of the 1D $p-$ wave superconducting system (see Eq.(\ref{ham1})). Phases $I$ and $II$ are topologically non-trivial while 
phase III is topologically trivial. Quenching paths A and B are shown in the phase diagram.}
\label{Fig1}
\end{figure}

The Hamiltonian of a one-dimensional $p$-wave superconducting system of spinless (or spin polarized) fermions with system size $N$ is given by
\ba
H &=&\sum_{j=1}^{N}[-w(a_j^{\dagger}a_{j+1}+a_{j+1}^{\dagger}a_j)+\Delta(a_ja_{j+1}+a_{j+1}^{\dagger}a_j^{\dagger})]\nonumber\\
&-& \sum_{j=1}^{N}\mu(a_j^{\dagger}a_j-1/2),
\label{ham1}
\ea
where $w$, $\Delta$ and $\mu$ denote nearest-neighbor hopping strength, superconducting gap and on-site chemical potential, respectively. 
We have considered spacing of the lattice as unity and also Planck constant, $\hbar =1$ throughout the paper.
The annihilation and creation operators $a_j$($a_j^{\dagger}$) obey the usual anticommutation relations $\{a_j,a_l\}=0$ and 
$\{a_j,a_l^{\dagger}\}=\de_{jl}$. The periodic boundary condition of the lattice ($a_{N+1}=a_1$) makes the Hamiltonian in Eq. (\ref{ham1}) 
translationally invariant and it 
can be then diagonalized in momentum basis, $a_k=\frac{1}{\sqrt{N}}\sum_{j=1}^N a_je^{-ikj}$.
This leads to a particle-hole symmetric dispersion given by
\be
E_k=\pm2w\sqrt{(\eta+\cos k)^2+\xi^2\sin^2k},
\label{spectrum1}
\ee
where we have introduced two relative parameters $\xi=\Delta/w$ and $\eta=\mu/2w$. 
The bulk energy gap (2$E_k$) vanishes at certain values of $\xi$ and $\eta$ for some specific $k$ modes. The phase diagram of the model with three 
distinct phases (denoted by I, II and III) is shown in Fig.~\ref{Fig1}. One observes that $\eta=\pm 1$ are two quantum critical lines with critical modes $k_c=\pi$ and $0$ (for which the energy gap vanishes), respectively, 
whereas for the critical line $\xi=0$  (with $\eta$ lies between $-1$ and $1$) with $k_c=\cos^{-1}(-\eta)$. 

We can represent the Hamiltonian (Eq.(\ref{ham1})) in terms of  Majorana fermion operators
\be
c_{2j-1}=a_j+a_j^{\dagger}\hspace{3mm}c_{2j}=\frac{1}{i}(a_j-a_j^{\dagger}).
\label{maj}
\ee
These Majorana operators are Hermitian and satisfy the relations $c_{2j-1}^2=c_{2j}^2=1$ and $\{c_j,c_l\}=2\de_{jl}$; these imply that  to one fermionic site $j$ there is two Majorana sites or equivalently,  a Majorana fermion can viewed as  occupying half of a state. 
Then the Eq. (\ref{ham1}) with an open boundary conditions can be re-written as
\ba
H &=&\frac{i}{2}\sum_{j=1}^{N-1}[(-w+\Delta)c_{2j-1}c_{2j+2}+(w+\Delta)c_{2j}c_{2j+1}]\nonumber\\
&-& \frac{i}{2}\sum_{j=1}^{N}\mu c_{2j-1}c_{2j}.
\label{ham2}
\ea

The zero energy modes of the Hamiltonian in Eq.(\ref{ham2}) distinguish different phases (I, II and III) in the phase diagram (see Fig.~\ref{Fig1}).
Phase I ($\xi>0$ and $|\eta|<1$) 
is one of the topological non-trivial phases, due to the presence of two isolated Majorana modes 
at the two edges of a open and long chain (see Fig.~\ref{Fig2}). It can be shown using a special condition i.e, for $\Delta=w$ and $\mu=0$, 
 when, as can be seen from Eq.(\ref{ham2}) that the Majorana operators $c_1$ and $c_{2N}$
 do not appear in the Hamiltonian leading to two unpaired zero energy 
Majorana fermions $c_1$ and $c_{2N}$ at the left and right end of the chain, respectively. Similarly, for the phase II ($\xi<0$ and $|\eta|<1$) two isolated 
Majorana modes, $c_2$ and $c_{2N-1}$, exist at the two ends of the chain. As a result, the ground state of the system is twofold degenerate with definite fermionic parity
for this two phases. (One can combine the two unpaired Majorana fermions to get a usual complex fermion, giving two degenerate ground states with fermionic occupation 
number $0$ or $1$.) We note that  the phases I and II are different with respect to the nature of edge modes in two ends of the chain. While Phase I hosts odd (even)
site edge Majorana modes at the left (right) of the open and long chain, it is the other way round in Phase II.
The phase III ($|\eta|>1$) is  topologically trivial with no edge Majorana modes.  This can be illustrated  considering the special limit $\Delta=w=0$, $\mu\neq0$, when all the Majorana fermions are connected pairwise for each fermionic lattice site and 
consequently there is no isolated edge states.  The  system has a topological invariant (TI) through which we can distinguish different phases;  this is known as
the winding number \ct{niu12,tong12} which takes values $-1$, $1$ and $0$ in  phases I, II and III, respectively.
One can also transform the Hamiltonian (Eq.(\ref{ham2})) to a spin-$1/2$ $XY$ model Hamiltonian with a transverse 
field by using the Jordan-Wigner transformation\ct{lieb61}.
\section{Quench dynamics and results}
\label{III}
In this section, we will study the survival probability of an edge Majorana when the superconducting chain is quenched from one phase to the other or
to the QCP separating two phases as shown in Fig.~\ref{Fig1}. 
 Immediately after the quench the quantum state is not an the eigenstate of the final Hamiltonian and hence evolves in time with time evolution
 dictated by the final Hamiltonian.
 
\begin{figure}[ht]
\begin{center}
\includegraphics[height=2.2in]{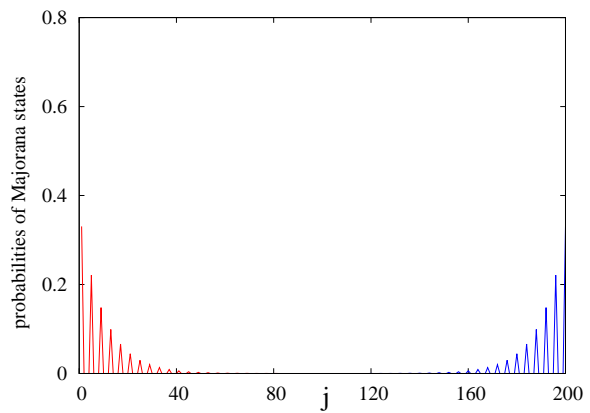}
\end{center}
\caption{(Color online) Two isolated Majorana states are localized in two edges of a $100$-site open Majorana chain in phase I ( $\xi=0.1$ and $\eta=0.0$). 
$j$ labels as Majorana sites $1,2,...200$.
One can see that here probability is non-zero only if $j$ is odd (even) for the left (right) end of Majorana chain.}
\label{Fig2}
\end{figure}

To illustrate the method to be employed here, we represent the Eq.\ref{ham2} in a generic quadratic form with Majorana operators
\be
H=\frac{i}{4}\sum_{j,k=1}^{2N}A_{jk}c_jc_k,
\label{ham4}
\ee
where $A$ is a real skew-symmetric $2N\times2N$ matrix and the $c_j$ are the Majorana operators;  the eigenvalues of $A$ appear in pairs
as $\pm i\epsilon_j$ 
($j$=$1,2$,...,$2N$). The zero eigenvalues of the above Hamiltonian will be even in number and these are called zero modes. For an open and long  chain the wave functions of zero energy modes can be made real and they 
have finite gap with respect to other eigenvalues. These properties lead to identification of the end modes as Majorana states\ct{thakurathi13}.
Two isolated edge Majorana states existing in the phase I is shown in Fig.~\ref{Fig2}; these are obtained by  diagonalizing the Hamiltonian in Eq.(\ref{ham4}) with 
given parameter values  to obtain the zero eigenvalues and constructing the appropriate real wave functions.
 
\begin{figure}[ht]
\begin{center}
\includegraphics[height=1.8in]{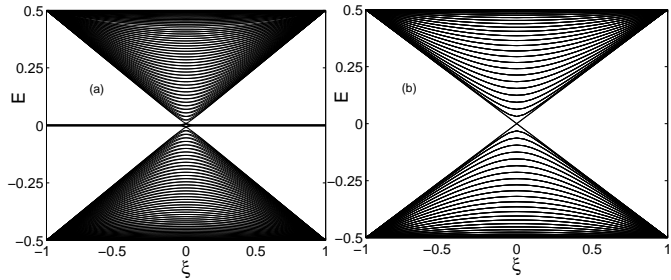}
\end{center}
\caption{Energy spectrum of the Majorana chain of system size $N=100$ as a function of $\xi$ with $w=1$ and $\eta=0$ using (a) open boundary condition (OBC) 
and (b) periodic boundary condition (PBC), respectively. Note that two zero-energy Majorana modes are present in case (a) but
not in case (b). We mention that the energy is scaled by a factor $1/4$  in Eq.(\ref{ham4}) in comparison to
Eq.(\ref{spectrum1}).}
\label{Fig3}
\end{figure}

 
Let us now analyze the  energy spectrums for the Majorana chain for open and and periodic boundary conditions  as shown 
in Fig.~\ref{Fig3}, for $\xi\in[-1,1]$ with $\eta=0$. We have set here $w=1$; a convention that would be followed for rest of the paper.
 We can clearly identify that the Hamiltonian in 
Eq.(\ref{ham4}) carries two zero energy modes in both the phases I and II for a open  chain; but periodic chain has no zero energy states and
hence  it 
is a purely non-topological system. 

We now quench rapidly a parameter of the Hamiltonian in Eq.(\ref{ham4}) considering different paths on the phase diagram (Fig.~\ref{Fig1}) 
at the instant $t=0$ and study the time evolution of a zero energy Majorana 
mode. Following a sudden quench the time evolved state of the Majorana at time $t$  can be  written as
\be
\ket{\psi_m(\xi,\eta,t)}=\sum_{n=1}^{2N}e^{-iE_nt}\ket{\Phi_n(\xi',\eta')}\langle\Phi_n(\xi',\eta')|\psi_m(\xi,\eta)\rangle,
\label{majmode}
\ee
where $|\psi_m(\xi,\eta)\rangle$ is an initial end Majorana state for the parameters $\xi$ and $\eta$, and $|\Phi_n(\xi',\eta')\rangle$ are the eigenstates of the 
final Hamiltonian with new parameters $\xi'$, $\eta'$ while  $E_n$'s are the eigenvalues of the final Hamiltonian. These are obtained  by diagonalizing the Hamiltonian in Eq.(\ref{ham4}) numerically for the initial and final parameter values.
To study the dynamics of the zero energy Majorana mode after a sudden quench, we now define the Majorana survival probability (MSP) $P_m(t)$ (which is infact the 
Loschmidt echo (LE)\ct{quan06,sharma12}) defined as
\be
P_m(t)=\Big|\sum_{n=1}^{2N}|\langle\psi_m(\xi,\eta)|\Phi_n(\xi',\eta')\rangle|^2e^{-iE_nt}\Big|^2,
\label{prob}
\ee
which we shall analyze  below for different quenching paths.

\subsection{Quenching along different paths}
\label{a}
\begin{figure}[ht]
\begin{center}
\includegraphics[height=2.4in]{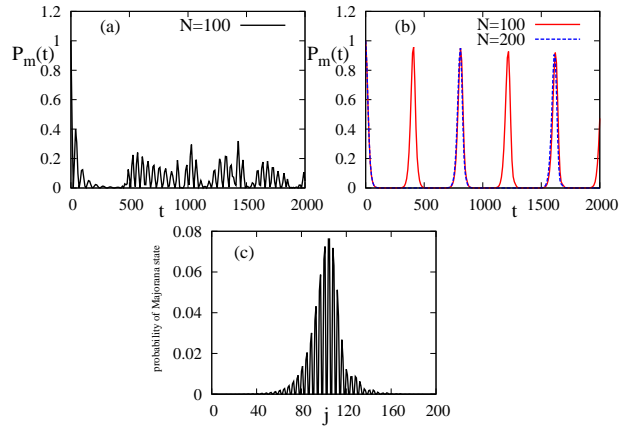}
\end{center}
\caption{(Color online) Majorana survival probability of a zero energy Majorana for different quench schemes along the path A and probability of Majorana after quench.
(a) MSP decays rapidly and does not revive significantly for quenching from phase I ($\xi=0.1$) to phase II ($\xi=-0.1$). (b) Quenching from phase I ($\xi=0.1$) to
the QCP ($\xi=0.0$), MSP shows nearly perfect collapse and revival with time $t$ and scales linearly with the system size $N$.
(c) Probability of the end Majorana 
at time $t=100$ after quench at the QCP with the Majorana site $j$. This shows that at this instant the probability of Majorana is
maximum around the center of the chain.}
\label{Fig4}
\end{figure}

We have considered a path A ($\eta=0$) in the Fig.~\ref{Fig1}. Along this path we change the parameter $\xi$ suddenly from phase I to II.
The probability defined in Eq.(\ref{prob}), decays rapidly with time and remains minimum having some noisy fluctuation of small amplitude (Fig.~\ref{Fig4}(a)) for first 
quench scheme (phase I to II). This signifies that the Majorana mode de-coheres with time, but not completely due to unitary evolution of the system.

\begin{figure}[ht]
\begin{center}
\includegraphics[height=1.6in]{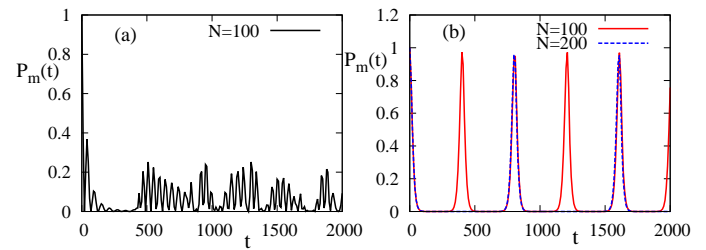}
\end{center}
\caption{(Color online) The probablities of an end mode ($\xi=1.0$ and $\eta=0.9$) after a quench (a) at a point ($\xi=1.0$ and $\eta=1.1$) of phase III and (b) at a 
QCP ($\xi=1.0$ and $\eta=1.0$) along the path B show same behavior as of the path A.}
\label{Fig5}
\end{figure}

Let us now focus on the most interesting situation which emerges when we quench the chain to the quantum critical point (QCP) ($\xi=0$, $\eta=0$); we find that the MSP of an edge  Majorana mode shows collapse and revival with time (Fig.~\ref{Fig4}$(b)$). 
This can be attributed to the nearly equi-spaced levels of the bulk close at $\xi=0$ (see Fig.~\ref{Fig3}); moreover there is bulk bandgap $E_g\sim 1/N$ 
as dynamical critical exponent associated with this QCP is unity.

The significant contribution to the summation of Eq.~(\ref{prob}) arises because of the overlap of the edge Majorana state with  of these (nearly equispaced)
  low-energy states of the bulk. Then the Eq.~(\ref{prob}) represents Fourier series of a periodic function 
with fundamental frequency as an energy difference of two consecutive equispaced levels i.e., proportional to $1/N$; this in turn implies
that the quasi period of revival should scale with the chain length $N$.
One can confirm this 
prediction from Fig.~\ref{Fig4}(b) where the time period of oscillation getting double when system size becomes twice of earlier. 
 These  observations lead to the conclusion that  when the chain quenched at the QCP separating two topological phases the edge Majorana mode of Fig.~\ref{Fig2}  oscillates between two ends 
of the chain (see Fig.~\ref{Fig4}(c)) with  the time period of this oscillation being proportional to the size of the spin chain. However, the amplitude
of revival decreases with time due to the coupling of the Majorana with bulk levels of higher energy as it oscillates back and forth between
the two edges.

To analyze this oscillation of the Majoranas further, we refer to the situation with
the periodic boundary condition; although in this case,  though the edge Majorana does not exist but  the structure of bulk energy levels remains the same.
We can then write  the Hamiltonian in Eq.(\ref{ham1}) as the direct product of decoupled  $2\times2$ matrices in each Fourier mode $k$; the 
bulk energy dispersion is given by the Eq.(\ref{spectrum1}); 
at the critical point $(\xi=0,\eta=0)$, the bulk energy gap vanishes for the critical mode $k_c=\pi/2$. Now, if we expand $E_k$ close to the critical mode $k_c=\pi/2$ 
the first non-vanishing term in $E_k$ is linear order in $k$ and second term will be proportional to $k^3$ (where we have rescaled $(k-k_c) \to k$). 
If one starts from the ground state in one of the phases which evolves  with the critical Hamiltonian, the modulus of the overlap between the initial
state and the time evolved state (or the Loschmidt echo) takes
a form:  $ {\cal L}(t)= \prod_{k>0} \left(1 - A_k \sin^2 (E_kt) \right)$, \ct{quan06,sharma12,happola12}, where $A_k$ is a sufficiently slowly varying
function of $k$ and each term has a periodicity $\pi/E_k$.  We now focus on the modes close to $k \to 0$ separated by $\Delta k =2\pi/N$; these
modes interfere constructively at  time instants given $\Delta E_k t = p\pi$, where $p$ is an integer; yielding revival time for these modes given by $t_k= \frac 1 {2} p N |\partial E_k/\partial k|$.
For a strictly linear dispersion (valid for modes with $k \to 0$), when $|\partial E_k/\partial k|$ is independent of $k$,  many modes add up constructively and one finds a pronounced revival of the overlap at intervals which scale with chain length $N$.
Noting that the  bulk energy difference, $\Delta E=E_n-E_{n-1}\approx 2\pi/N$ (with $w=1$), between two consecutive low-energy states close to $\xi=0$. As a 
	consequence, the time-period of revival is given by 
	\be
	T= \frac{2\pi}{\Delta E}\propto N.
	\label{time_period}
	\ee
However, the amplitude of the revival decay with time; if  one considers the next higher order non-linear term in the expansion of $E_k \sim k^3$,
these modes do not interfere constructively and there is an eventual decay  in the amplitude as the Majorana oscillates back and forth between
the edges; asymptotically the MSP saturates at a finite value for a finite size system.

Let us now choose a path B ($\xi=1.0$) that crosses or terminates at the quantum critical line separating topological (I) and non-topological (III) phases. 
The quenching results (Figs.~\ref{Fig5}(a),(b)) for the both cases i.e., quenching from 
phase I to phase III and upto QCP are exactly same as of path A which can be explained using a similar set of arguments.

\begin{figure}[ht]
\begin{center}
\includegraphics[height=1.6in]{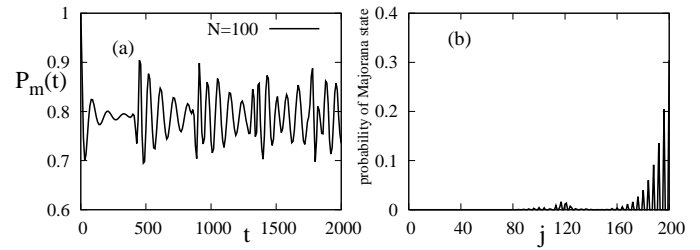}
\end{center}
\caption{System is suddenly quenched within the same phase (I) from ($\xi=0.2$ and $\eta=0.0$) to ($\xi=0.1$ and $\eta=0.0$) and MSP becomes (a) rapidly fluctuating function 
of time $t$ with a mean value of nearly $0.8$ and (b) probability of the end Majorana at $t=100$, with the Majorana sites $j$ after quench.}
\label{Fig6}
\end{figure}


\subsection{Quenching within the same phase}
\label{b}
In this case the parameters in the Hamiltonian in Eq.(\ref{ham2}) is quenched within the same phase (phase I). Due to quenching MSP decays, but it does  not collapse at zero because bulk 
of the system has a pronounced gap and the system does not go to the gapless QCP. Also, there is no perfect overlap   due to the unequal spacing of energy levels. The MSP rapidly fluctuates  haphazardly with a mean 
value close to unity (see Fig.~\ref{Fig6}(a)).  We conclude that the zero energy Majorana mode is mainly  localized at the end of the chain  after this quenching (see Fig.~\ref{Fig6}(b)).

\section{Quenching in the presence of a next-nearest-neighbor interaction}
\label{IV}
In this section we introduce the Hamiltonian for 1-D $p$-wave superconductor with next-nearest-neighbor hopping and superconducting pairing which 
lead to  two isolated edge Majorana modes at each end of the chain with open boundary condition. Our interest is to study how does the presence of two Majorana modes at each end  shows up in the MSP following a sudden quenching.
The Hamiltonian for such a 1-D chain is given by \ct{wdegottardi13}, 
\ba
H &=&\sum_{j=1}^{N-1}[-w(a_j^{\dagger}a_{j+2}+a_{j+2}^{\dagger}a_j)+\Delta(a_ja_{j+2}+a_{j+2}^{\dagger}a_j^{\dagger})]\nonumber\\
&-& \sum_{j=1}^{N}\mu(a_j^{\dagger}a_j-1/2).
\label{ham5}
\ea
In this case $w$ and $\Delta$ are the next-nearest-neighbor hopping strength and superconducting pairing term, respectively. In an identical
fashion  to that used for the nearest neighbor chain
(see Eq.(\ref{spectrum1})), one can find out the energy spectrum for the Hamiltonian (Eq.(\ref{ham5})), given by
\be
E_k=\pm2w\sqrt{(\eta+\cos 2k)^2+\xi^2\sin^22k}.
\label{spectrum2}
\ee
Comparing with Eq.~(\ref{spectrum1}), we note that due to the next nearest neighbor interaction, there is a shift of wave vector $k \to 2k$ in the spectrum.

To describe the topological properties of this system,  we represent the Hamiltonian in Eq.(\ref{ham5}) using Majorana fermions with an open
boundary  condition, which now takes the form
\ba
H &=&\frac{i}{2}\sum_{j=1}^{N-1}[(-w+\Delta)c_{2j-1}c_{2j+4}+(w+\Delta)c_{2j}c_{2j+3}]\nonumber\\
&-& \frac{i}{2}\sum_{j=1}^{N}\mu c_{2j-1}c_{2j}.
\label{ham6}
\ea

\begin{figure}[ht]
\begin{center}
\includegraphics[height=2.2in]{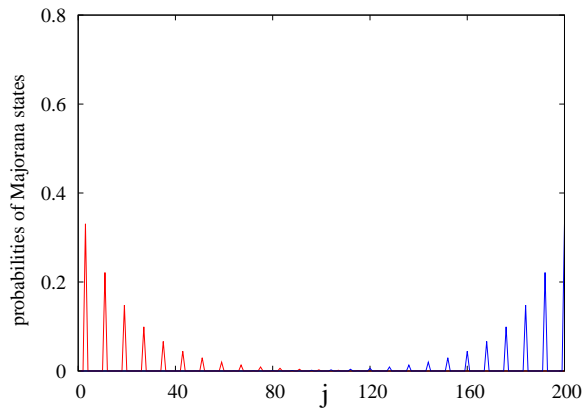}
\end{center}
\caption{(Color online) Two zero energy Majorana modes of a system Hamiltonian defined in Eq.(\ref{ham6})
exist in two edges of a $100$-site open Majorana chain in phase I ($\xi=0.1$ and $\eta=0.0$). 
Similarly to Fig.~\ref{Fig2}, here also $j$ labels as Majorana sites $1,2,...200$. 
The probabilities are non-zero only if $j$ is odd (even) for the left (right) end of Majorana chain.}
\label{Fig7}
\end{figure}

\begin{figure}[ht]
\begin{center}
\includegraphics[height=2.2in]{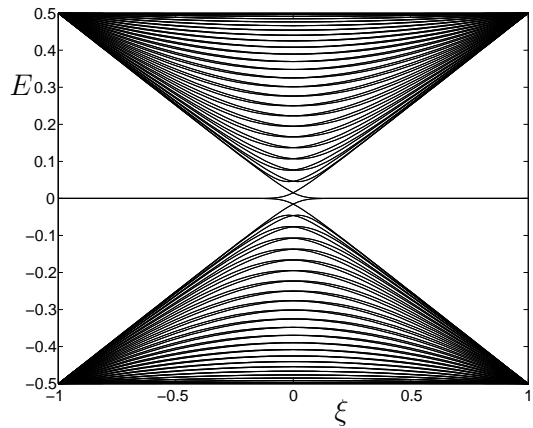}
\end{center}
\caption{Energy spectrum of the Majorana chain (with only next-nearest-neighbor hopping and interaction as defined in Eq.({\ref{ham5}})) as a function of $\xi$ with $\eta=0$ 
and $N=100$ sites for open boundary condition. In this case also a energy scale difference exists (see caption of Fig.~\ref{Fig3}).}
\label{Fig8}
\end{figure}

The phase diagram also consists of two topological phases (I, II) and a non-topological phase  (III); however,  the  phases I and II have two isolated Majorana 
fermions in each end of the chain. We can illustrate this again using 
an extreme limit;  for $\xi=1$ and $\eta=0$, the  chain carries two zero-energy Majorana modes $c_1$ and $c_3$ in left end and also two modes 
$c_{2N-2}$ and $c_{2N}$ in right end. On the other hand, for $\xi=-1$ and $\eta=0$ lying in phase II, there are two zero-energy Majorana modes $c_2$ and $c_4$ in the 
left end and other two modes $c_{2N-3}$ and $c_{2N-1}$ in the right end of the open and long chain. In the Fig.~\ref{Fig7} we have shown one set of equilibrium edge Majorana modes 
for phase I in two ends of the chain obtained via diagonalizing the Hamiltonian (\ref{ham4}) for the present case. We notice that here the peaks of probabilities occur in double difference in Majorana sites compare to earlier case (see Fig.~\ref{Fig2}). 
This can be attributed to the next-nearest-neighbor hopping and superconducting interaction.
From Eq.\ref{spectrum2} we can see that there 
are two critical lines $\eta=\pm 1$ with critical modes $k_c=\pi/2$ and $0$, respectively. Another critical line, $\xi=0$ (separating phases I and II) exists 
with critical mode $k_c=\frac{1}{2}\cos^{-1}(-\eta)$.

\begin{figure}[ht]
\begin{center}
\includegraphics[height=2.4in]{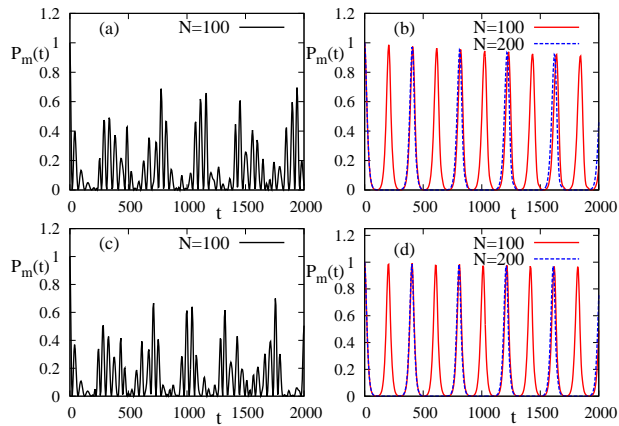}
\end{center}
\caption{(Color online) $P_m(t)$ of a right end zero energy Majorana mode after quenching along different paths. (a) MSP decays rapidly and stays minimum with some 
noisy fluctuation of small amplitude when the system (see Eq.(\ref{ham5})) is quenched from phase I ($\xi=0.1$) to the phase II ($\xi=-0.1$)
 along the path A. (b) For quenching to QCP $P_m(t)$ is nearly perfect oscillatory function of time $t$ with a interesting fact that its time period becomes 
half of the earlier case (see Fig.~\ref{Fig4}) and scales linearly with the system size $N$. 
(c) Time variation of MSP with $\xi=1.0$ and $\eta=0.9$ following a quench to a point ($\xi=1.0$ and $\eta=1.1$) and (d) the QCP ($\xi=1.0$ and $\eta=1.0$)
 along the  path B show same behavior as of the path A($\eta=0$).}
\label{Fig9}
\end{figure}

In this case also we have measured MSP for one of the edge Majoranas at one end  following a sudden change of the parameter in the Hamiltonian (Eq.(\ref{ham6})) considering the same paths as in Sec.~\ref{a}.
One can see in the Fig.~\ref{Fig9}(a) as usual rapid decay of the probability (introduced in Eq.(\ref{prob})) with time having some noisy fluctuation of small 
amplitude when the system is quenched from phase I to II along the path A ($\eta=0$). Now, we concentrate on the Fig.~\ref{Fig9}(b) which shows a noticeable change in the 
time period of collapse and revival of the probability in comparison   to Fig.~\ref{Fig4}(b) following a quench to QCP ($\xi=0$, $\eta=0$) from phase I the time period of oscillation becomes half of the earlier case (see Fig.~\ref{Fig4}(b)) for equal system size. This can be explained using the bulk energy spectrum 
of the system. We have discussed in Sec.\ref{a} that this perfect collapse and revival of MSP is a consequence of nearly equi-spaced levels of the bulk spectrum 
close to QCP ($\xi=0$, $\eta=0$).  
In the present case energy spacings close to QCP get doubled compared to the nearest neighbor 1-D chain for the same system size. 
This leads to energy difference between two consecutive equispaced levels is proportional to $2/N$. 
The doubling of spacings between two consecutive energy levels close to QCP can also be seen comparing  Fig.~\ref{Fig3} 
to Fig.~\ref{Fig8}. This in fact explains the reduction in the time period of collapse and revival of MSP in the present situation.
These observations imply that when this system is quenched suddenly to QCP separating two topological phases the edge Majorana takes half of the time
in comparison to nearest-neighbor case (see Fig.~\ref{Fig4}) for a full oscillation between two edges of the chain. In this case the end Majorana enters in the bulk with hopping to the next-nearest-neighboring sites and thereby avoiding subsequent sites.


We have also checked  that the MSP for the same quenching schedules with the path B ($\xi=1.0$) and the results come out to be same as of path $\eta=0$ 
(see Figs.~\ref{Fig9}(c),(d)). We can argue similar to above to explain the doubling of the frequency in collapse and revival of MSP.

\section{Quenching between a phase with two Majorana modes to a phase with one Majorana mode}
\label{V}
In this section, we shall label the Hamiltonians in Eqs.\ref{ham1} and \ref{ham5} as $H_1$ and $H_2$, respectively (both with open boundary conditions). Now we consider a time-dependent 
Hamiltonian as
\be
H(t)=\Theta(t)H_1+[1-\Theta(t)]H_2
\label{hamf}
\ee
where $\Theta(t)$ is a function whose value is $0$ for $t<0$ and $1$ for $t>0$, which implies that at $t=0$  Hamiltonian describing the system suddenly changes from  
$H_2$ to $H_1$ with the hopping amplitude and the superconducting gap being identical (in the topological phase I) in both the cases while the chemical 
potential $\mu$ set equal to zero. We recall that in the equilibrium the Hamiltonian $H_1$ has two Majoranas $c_1$ and $c_{2N}$ in the two ends of the chain, and 
similarly the Hamiltonian $H_2$ has a pair of Majoranas $c_1$ and $c_3$ in left and also $c_{2N-2}$ and $c_{2N}$ in right end.

\begin{figure}[ht]
\begin{center}
\includegraphics[height=1.6in]{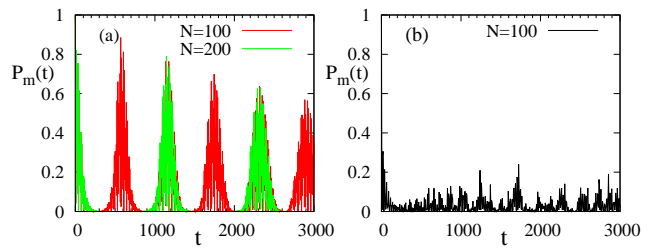}
\end{center}
\caption{System changes suddenly from a phase with two Majorana fermions to a phase which contains only one Majorana mode at each end of the chain at time $t=0$. We have 
set here the parameter values $\xi=0.1$ and $\eta=0.0$. 
 (a) MSP of a left end ($c_3$) Majorana mode shows collapse and revival with time, but at each revival there are fluctuations and also the peaks of revivals decrease
rapidly. (b) While on the other hand when the system is quenched reversely the Majorana ($c_1$) decoheres rapidly with time (with no prominet
revival) and fluctuates around zero mean.}
\label{Fig10}
\end{figure}

Our purpose here is to study the survival probability of one of the edge Majoranas residing at the left end when the Hamiltonian is changed suddenly from $H_2$ to $H_1$; 
result is presented in Fig.~\ref{Fig10} where we find oscillation in the survival probability. 
Referring to the situation depicted in Fig.~\ref{Fig10}(a), we note that the survival probability of $c_3$  oscillate as a function of time even though the $c_3$ Majorana should not 
exist in the final phase of the Hamiltonian in Eq.~(\ref{hamf}) and eventually the survival probability vanishes in the asymptotic limit. Remarkably,
the period of oscillation is again found to scale with $N$.
We have also probed the reverse situation, when the Hamiltonian is changed from $H_1$ to $H_2$, where we find a rapid decoherence
of the Majorana $c_1$ as shown in the right panel of Fig.~(\ref{Fig10}).

\begin{figure}[ht]
\begin{center}
\includegraphics[height=1.6in]{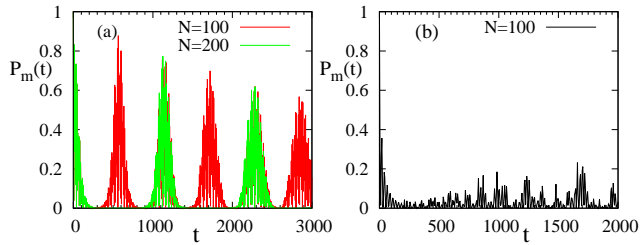}
\end{center}
\caption{Initially, the Hamiltonian $H_2$ is fixed at the parameter 
values $\xi=0.1$ and $\eta=0.0$ and then it is quenched to the critical point ($\xi=0.0$ and $\eta=0.0$) of the Hamiltonian $H_1$ at $t=0$,
 (a) MSP of a left end ($c_3$) Majorana mode shows collapse and revival with time quite similar to Fig.~\ref{Fig10}(a); the amplitude of the
 revival  decays rapidly if the initial value of $\Delta$ is larger, i,e., the Hamiltonian $H_2$ is deep in the topological phase away from
 the QCP.
 (b) There is a complete loss of coherence  when the system is quenched reversely.}
\label{Fig11}
\end{figure}

Remarkably, we find  similar oscillations are obtained by changing $H_2$ to $H_1$ when the latter is kept at the critical point with $\xi=0$ (see Fig.~\ref{Fig11}).
There is no such oscillation in the reverse case.

Oscillations obseved in the left panels of Figs.~\ref{Fig10} and \ref{Fig11} imply that apparently the system behaves in such a way that it is quenched to critical point of the final Hamiltonian 
eventhough there is a small $\Delta$; this is a surprising and remarkable result.

\section{Conclusions}
\label{VI}
In this paper, we report the possibility of survival of an edge Majorana  following a sudden quenching
of the p-wave superconducting chain. Analyzing the MSP, we show that when the chain quenched at the QCP separating the two topological phases (or separating one topological phase and
the trivial phase)  the edge Majorana mode   oscillates between two ends 
of the chain  with  the time period of this oscillation being proportional to the size of the spin chain. 
This phenomena can be attributed to nearly equispaced nature of the bulk modes and linearity
of the spectrum at the QCP. However, the amplitude of oscillation eventually decreases as
a function of time when the bulk modes (not equally-spaced) start  contributing  to the MSP. 
 We here also consider the effect of next-nearest-neighbor hopping and supeconducting 
pairing in MSP. In this case the frequency of oscillation of MSP is doubled when the system is quenched to a QCP. Finally, we have
looked at the instantaneous quenching between Hamiltonian $H_2$ (with two Majoranas at each end) and $H_1$ (which is supposed
to have one edge Majorana) when the latter is critical as well as off-critical; in both the cases, there are prominent revivals of
the survival probability with the time period scaling with the system size which is a remarkable result, not apparently obvious. 
There is also complete detraction of coherence when one performs a quench from $H_1$ to $H_2$.

\begin{center}
\bf{Acknowledgements}
\end{center}
We sincerely  thank Aavishkar Patel for helpful discussions and very useful comments. We also acknowledge useful discussions with Uma Divakaran, Tanay Nag and Shraddha Sharma.  AR thanks IIT Kanpur for hospitality during a part of this work.

\end{document}